\pdfoutput=1
\documentclass[11pt]{article}

\usepackage[]{ACL2023}

\usepackage{times}
\usepackage{latexsym}

\usepackage[T1]{fontenc}
\usepackage[utf8]{inputenc}

\usepackage{microtype}

\usepackage{inconsolata}

\usepackage{xspace} 
\usepackage{graphicx}
\usepackage{textcomp}
\usepackage{multirow}
\usepackage{colortbl}
\usepackage{hyperref}
\usepackage{hhline}
\usepackage{bbding}
\usepackage{makecell}
\usepackage{color}
\usepackage{xcolor}
\usepackage{arydshln}
\usepackage{booktabs}
\usepackage{amssymb}
\usepackage{vcell}
\usepackage{subcaption}

\newcommand{\nlcode}{NL2Code\xspace}

\newcommand{\passk}[0]{${\rm{pass}}@{k}$ }

\title{Large Language Models Meet \nlcode: A Survey}

\author{Daoguang Zan$^{1,2}$\thanks{~~This work was done before October 2022 when the author, Daoguang Zan, was an intern at Microsoft Research Asia.}, Bei Chen$^3$, Fengji Zhang$^{3}$, Dianjie Lu$^4$, Bingchao Wu$^{1}$,\\
\textbf{Bei Guan$^{5}$, Yongji Wang$^{5}$, Jian-Guang Lou$^3$} \\
  $^1$Cooperative Innovation Center, Institute of Software, Chinese Academy of Sciences \\
  $^2$University of Chinese Academy of Sciences;
  $^3$Microsoft Research Asia;
  $^4$Shandong Normal University \\
  $^5$Integrative Innovation Center, Institute of Software, Chinese Academy of Sciences \\
  \texttt{\{daoguang@, bingchao2017@, guanbei@, ywang@itechs.\}iscas.ac.cn}; \\
  \texttt{\{beichen, v-fengjzhang, jlou\}@microsoft.com}; \texttt{Ludianjie@sdnu.edu.cn} \\
}

\begin{document}
\maketitle
\begin{abstract}
The task of generating code from a natural language description, or \nlcode, is considered a pressing and significant challenge in code intelligence. Thanks to the rapid development of pre-training techniques, surging large language models are being proposed for code, sparking the advances in \nlcode. To facilitate further research and applications in this field, in this paper, we present a comprehensive survey of $27$ existing large language models for \nlcode, and also review benchmarks and metrics. We provide an intuitive comparison of all existing models on the HumanEval benchmark. Through in-depth observation and analysis, we provide some insights and conclude that the key factors contributing to the success of large language models for \nlcode are "Large Size, Premium Data, Expert Tuning". In addition, we discuss challenges and opportunities regarding the gap between models and humans. We also create a website \url{https://nl2code.github.io} to track the latest progress through crowd-sourcing. To the best of our knowledge, this is the first survey of large language models for \nlcode, and we believe it will contribute to the ongoing development of the field.

\end{abstract}

\section{Introduction} \label{sec:introduction}
Is it possible for novice programmers, even those without any programming experience, to create software simply by describing their requirements in natural language? This is a long-standing fascinating question, which poses challenges to research areas like software engineering, programming language, and artificial intelligence. 
Realizing this scenario would have an unprecedented impact on our lives, education, economy, and labour market, as it would change the centralized software development and operation paradigm. Due to its promising and intriguing future, natural-language-to-code (\nlcode) has been proposed as a research task that has attracted widespread interest in both academia and industry, with the goal of generating code from natural language descriptions.

Early studies on \nlcode were mainly based on heuristic rules or expert systems, such as probabilistic grammar-based methods~\cite{pg-3,pg-4,pg-5} and those focusing on domain-specific languages~\cite{dsl-6,dsl-1,dsl-2}, which are inflexible and not scalable. Other studies utilized static language models, like n-gram~\cite{ngram-2,ngram-3,ngram-1} and Hidden Markov~\cite{hidden-markov}, which have sparse vector representations and cannot model long-term dependencies.
Subsequently, neural networks, including CNN~\cite{cnn-2,cnn-1}, RNN~\cite{rnn-lstm-1,rnn-2}, and LSTM~\cite{lstm-2,lstm-4}, were employed to model the relationship between NL and code. In 2017, the Transformer~\cite{transformer} model was introduced for machine translation and later applied to the \nlcode task~\cite{transformer-1,transformer-2}.
However, these deep learning models require a significant amount of labelled pairs of NL and code for training, and have limited capabilities for the \nlcode task.

Recently, a growing number of large language models (LLMs) with Transformer architecture have been trained on large-scale unlabelled code corpus. These models have the ability to generate code in a zero-shot manner and have achieved impressive results in the \nlcode task. As a milestone, Codex~\cite{codex} has shown that an LLM with $12$ billion parameters is able to solve $72.31$\% of challenging Python programming problems created by humans.
More encouragingly, Codex has been used to power a commercial product\footnote{\url{https://github.com/features/copilot}} and improve coding efficiency in practice~\cite{efficiency-1,efficiency-2}.
Following Codex's success, various LLMs for the \nlcode task have emerged, with model sizes ranging from millions to billions of parameters.  
Examples include AlphaCode~\cite{alphacode}, which aims to solve competitive-level programming problems, and
InCoder~\cite{incoder}, which supports filling code in arbitrary positions using bidirectional contexts.
Other models such as CodeGen~\cite{codegen}, PaLM-Coder~\cite{palm}, PanGu-Coder~\cite{pangu-coder}, CodeGeeX~\cite{codegeex}, and SantaCoder~\cite{santacoder} have also gained great attention. 
As the model size increases, LLMs have been shown to exhibit some emergent capabilities such as human-like programming and debugging~\cite{bug-location-fix,bug-1,kang2023explainable}.

\begin{figure}[t]
    \small
    \centering
    \includegraphics[width=0.95\linewidth]{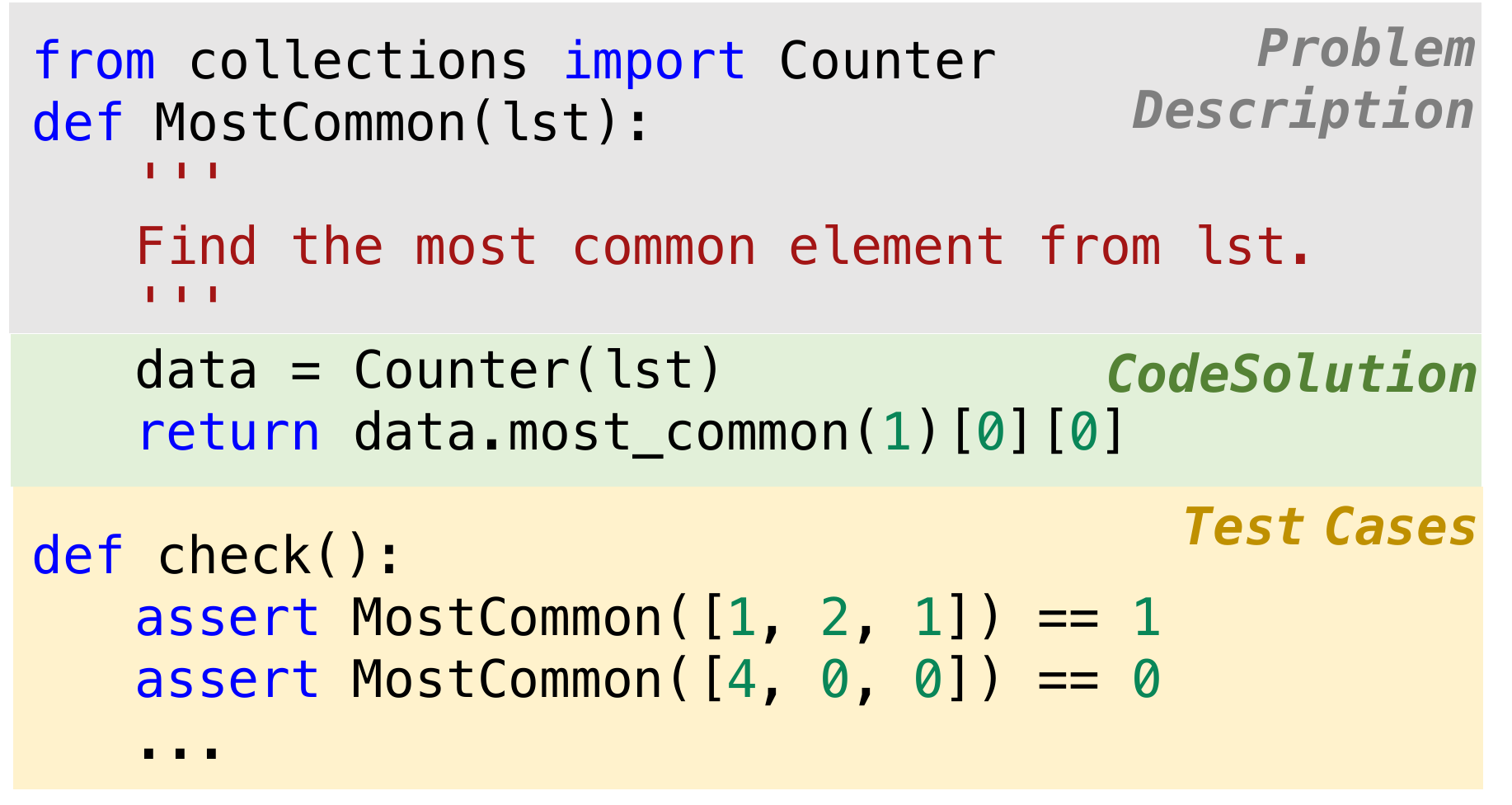}
   \caption{A simple example of the \nlcode task. The code blocks marked in grey, green, and yellow represent the natural language problem description, the predicted code solution, and the test cases, respectively.}
    \label{fig:programming-problem-example}
\end{figure}

Large language models have kindled hope for the \nlcode task due to their impressive power and potential value. Despite the significant progress, there are still numerous challenges and opportunities, calling for more advanced and innovative future work. Currently, considering the variety of techniques and applications, there is a growing need for a comprehensive survey to provide a systematic overview of this field and identify critical challenges. To this end, in this paper, we carefully investigate $27$ advanced LLMs for \nlcode (\S\ref{sec:models}), and also review benchmarks and metrics (\S\ref{sec:benchmarks}). 
We conduct an intuitive comparison of all the existing LLMs on the HumanEval benchmark, perform a thorough analysis, and eventually attribute the success of these LLMs to "\emph{Large Size, Premium Data, Expert Tuning}" (\S\ref{sec:analysis}). This means large model and data size, high-quality training data and expert hyper-parameter tuning. 
We also discuss the challenges and opportunities regarding the ability gap between LLMs and Humans (\S\ref{sec:challenges-and-future-works}). In addition, we have built a website \url{https://nl2code.github.io} to keep track of the latest progress and support crowd-sourcing updates. To the best of our knowledge, this is the first survey of LLMs for \nlcode\footnote{We summarize the related surveys in Appendix \ref{apx:related-survey}.}, and we hope it will contribute to the ongoing development of this exciting field.

\begin{table}[t]
\centering
\scalebox{0.75}{
\rotatebox{0}{
\begin{tabular}{llclll} 
\toprule
\textbf{Model} & \textbf{Size}             & \textbf{L.} & \textbf{A.} & \textbf{H.}  & \textbf{P.}\\ 
\hline
\multicolumn{6}{c}{\textbf{\textit{Decoder}}}               \\ 
\hline
GPT-C~\citeyearpar{gpt-c}          & $366$M                  & $24$  & $16$  & $1,024$     & {\color[rgb]{1,0,0}\texttimes}              \\
CodeGPT~\citeyearpar{codexglue}        & $124$M              & $12$  & $12$  & $768$         & {\color[rgb]{0,0.5,0}\checkmark}            \\ 
GPT-Neo~\citeyearpar{gpt-neo}        & $125$M\textasciitilde{}$2.7$B               & $32$  & $20$  & $2,560$    & {\color[rgb]{0,0.5,0}\checkmark}              \\
GPT-J~\citeyearpar{gpt-j}          & $6$B                    & $28$  & $16$  & $4,096$            & {\color[rgb]{0,0.5,0}\checkmark}              \\
Codex~\citeyearpar{codex}          & $12$M\textasciitilde{}$12$B                   & $40$  & $40$  & $5,140$      & {\color[rgb]{1,0,0}\texttimes}              \\
GPT-CC~\citeyearpar{codeclippy}     & $125$M\textasciitilde{}$1.3$B                 & $24$  & $16$  & $2,048$  & {\color[rgb]{0,0.5,0}\checkmark}              \\
CodeParrot~\citeyearpar{codeparrot}     & $110$M\textasciitilde{}$1.5$B             & $48$  & $25$  & $1,600$    & {\color[rgb]{0,0.5,0}\checkmark}              \\
LaMDA~\citeyearpar{lamda}          & $2$B\textasciitilde{}$137$B                 & $64$  & $128$  & $8,192$  & {\color[rgb]{1,0,0}\texttimes}              \\
PolyCoder~\citeyearpar{polycoder}      & $160$M\textasciitilde{}$2.7$B             & $32$  & $32$  & $2,560$ & {\color[rgb]{0,0.5,0}\checkmark}              \\
CodeGen~\citeyearpar{codegen}        & $350$M\textasciitilde{}$16.1$B               & $34$  & $24$  & $6,144$       & {\color[rgb]{0,0.5,0}\checkmark}              \\
InCoder~\citeyearpar{incoder}        & $1.3$B\textasciitilde{}$6.7$B                & $32$  & $32$  & $4,096$ & {\color[rgb]{0,0.5,0}\checkmark}              \\
GPT-NeoX~\citeyearpar{gpt-neox-20b}       & $20$B            & $44$  & $64$  & $6,144$           & {\color[rgb]{0,0.5,0}\checkmark}              \\
PaLM-Coder~\citeyearpar{palm}     & $8$B\textasciitilde{}$540$B                   & $118$  & $48$  & $18,432$    & {\color[rgb]{1,0,0}\texttimes}              \\
PanGu-Coder~\citeyearpar{pangu-coder}    & $317$M\textasciitilde{}$2.6$B            & $32$  & $32$  & $2,560$    & {\color[rgb]{1,0,0}\texttimes}              \\
FIM~\citeyearpar{fim}        & $50$M\textasciitilde{}$6.9$B                        & $32$  & $32$  & $4,096$ & {\color[rgb]{1,0,0}\texttimes}              \\
PyCodeGPT~\citeyearpar{cert}      & $110$M                   & $12$  & $12$  & $768$   & {\color[rgb]{0,0.5,0}\checkmark}              \\
CodeGeeX~\citeyearpar{codegeex}       & $13$B                & $39$  & $40$  & $5,120$       & {\color[rgb]{0,0.5,0}\checkmark}              \\
BLOOM~\citeyearpar{bloom}          & $560$M\textasciitilde{}$176$B                 & $70$  & $112$  & $14,336$  & {\color[rgb]{0,0.5,0}\checkmark}              \\
SantaCoder~\citeyearpar{santacoder}     & $1.1$B            & $24$  & $16$  & $2,048$   & {\color[rgb]{0,0.5,0}\checkmark}              \\ 
\hline
\multicolumn{6}{c}{\textbf{\textit{Encoder-Decoder}}}       \\ 
\hline
PyMT5~\citeyearpar{pymt5}          & $374$M                 & $12$  & $16$  & $1,472$     & {\color[rgb]{1,0,0}\texttimes}              \\
PLBART~\citeyearpar{plbart}         & $140$M\textasciitilde{}$406$M                & $24$  & $16$  & $1,024$ & {\color[rgb]{0,0.5,0}\checkmark}              \\
CodeT5~\citeyearpar{codet5}         & $60$M\textasciitilde{}$770$M                & $48$  & $16$  & $1,024$  & {\color[rgb]{0,0.5,0}\checkmark}              \\
JuPyT5~\citeyearpar{jupyt5}         & $350$M                & $12$  & $16$  & $1,472$     & {\color[rgb]{1,0,0}\texttimes}              \\
AlphaCode~\citeyearpar{alphacode}      & $284$M\textasciitilde{}$41.1$B              & $64$  & $128$  & $6,144$ & {\color[rgb]{1,0,0}\texttimes}              \\
CodeRL~\citeyearpar{coderl}         & $770$M                & $48$  & $16$  & $1,024$      & {\color[rgb]{0,0.5,0}\checkmark}              \\
CodeT5Mix~\citeyearpar{codet5mix}      & $220$M\textasciitilde{}$770$M             & $48$  & $16$  & $1,024$   & {\color[rgb]{0,0.5,0}\checkmark}              \\
ERNIE-Code~\citeyearpar{ernie-code}     & $560$M            & $24$  & $12$  & $768$          & {\color[rgb]{0,0.5,0}\checkmark}              \\
\bottomrule
\end{tabular}
}
}
\caption{Summary of $27$ existing LLMs for \nlcode. We show L. (number of layers), A. (number of attention heads), H. (hidden dimensions), and P. (model weights public or not) for the largest size version of each model.
Note that some models, such as GPT-Neo, GPT-J, LaMDA, GPT-NeoX, FIM, and BLOOM, are not exclusively trained for code.
}
\label{tab:models-size-pub}
\end{table}

\begin{figure*}[t]
    \small
    \centering
    \includegraphics[width=1.0\linewidth]{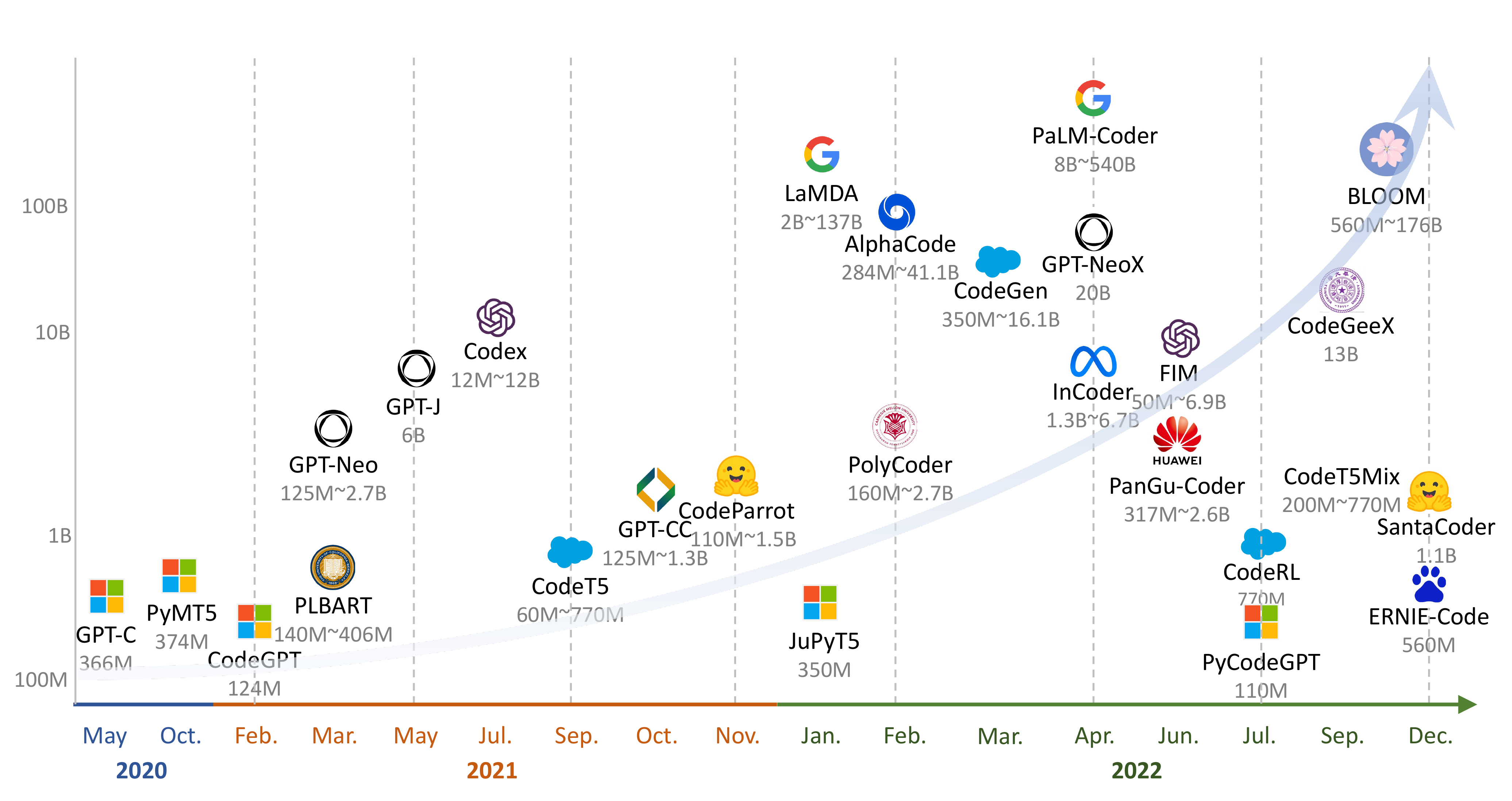}
    \caption{The timeline of LLMs for \nlcode, with only the largest model sizes plotted for visual clarity.}
    \label{fig:model-time-params}
\end{figure*}

\section{Large Language Models for \nlcode} \label{sec:models}
Given a natural language problem description, the \nlcode task aims to automatically generate the demanded code. To illustrate this task visually, we provide a Python programming problem as an example in Figure~\ref{fig:programming-problem-example}, while different \nlcode benchmarks may vary in terms of language or problem domain.
Existing large language models for the \nlcode task are usually based on Transformer~\cite{transformer} and are trained on a large-scale code related unlabelled corpus. 
For better code generation performance, most LLMs, no matter encoder-decoder or decoder-only models, employ the causal language modeling objective for training, which is to predict the token following a sequence of tokens.
During inference, an LLM can tackle \nlcode problems in a zero-shot manner without fine-tuning its parameters. There are also studies employing few-shot~\cite{mbpp} or in-context learning~\cite{codegen} to further boost the performance.

We conduct a comprehensive investigation of $27$ representative LLMs for the \nlcode task. Details of each model are summarized in Table~\ref{tab:models-size-pub}, where models vary in architecture, size, and accessibility. For better visualization, we present these models in chronological order in Figure~\ref{fig:model-time-params}, plotting the largest model sizes. One trend observed is that these large language models are consistently growing in size as the research field advances. Additionally, the decoder-only architecture is favoured for pre-trained models with larger sizes.

Early works, such as GPT-C~\cite{gpt-c}, PyMT5~\cite{pymt5}, and PLBART~\cite{plbart}, have relatively small numbers of parameters and do not demonstrate strong capabilities in zero-shot code generation. Conversely, large-scale models such as GPT-Neo~\cite{gpt-neo} and GPT-J~\cite{gpt-j}, despite their billion-level parameter scale, have been found to have limited power in the \nlcode task due to the small amount of code in their training corpus. 
Recently, a number of powerful LLMs have been proposed for \nlcode, such as Codex~\cite{codex}, AlphaCode~\cite{alphacode}, and PaLM-Coder~\cite{palm}, which possess massive parameter scales and high-quality training corpus with code. While they show surprisingly good performance on \nlcode, most of them are not readily accessible. 
At present, a number of excellent open-source models have also been proposed, including CodeParrot~\cite{codeparrot}, PolyCoder~\cite{polycoder}, GPT-NeoX~\cite{gpt-neox-20b}, and SantaCoder~\cite{santacoder}, which contribute to the thriving of LLMs for \nlcode. 
Besides, recent studies have proposed various approaches to address specific \nlcode scenarios. For example, JuPyT5~\cite{jupyt5} is designed to work within Jupyter Notebooks, while ERNIE-Code~\cite{ernie-code}, CodeGeeX~\cite{codegeex}, and BLOOM~\cite{bloom} are trained to support multiple natural or programming languages. 
Additionally, InCoder~\cite{incoder}, FIM~\cite{fim}, and SantaCoder~\cite{santacoder} not only support left-to-right code prediction, but also allow for infilling arbitrary regions of code. 
As LLMs for \nlcode are evolving rapidly, we created a website to keep up-to-date with the latest advances by crowd-sourcing. Details of the website can be found in Appendix~\ref{apx:website}.

These models are not only attractive in academia~\cite{codex,codegen,alphacode}, but also applied in real-world products to improve programming efficiency~\cite{efficiency-1,efficiency-2}. One example is GitHub and OpenAI's Copilot, a programming assistance tool that utilizes Codex to provide real-time code suggestions. Other notable products include CodeGeeX\footnote{\url{https://keg.cs.tsinghua.edu.cn/codegeex}} and CodeWhisperer\footnote{\url{https://aws.amazon.com/cn/codewhisperer}}. A summary of $10$ products can be found in Appendix Table~\ref{tab:products}. Recent studies~\cite{copilot-1,copilot-2,copilot-3} have shown that these products can provide helpful recommendations, while they also introduce minor bugs that can cause issues for users. There is still room for improvement before LLMs can be fully practical and capable of coding like humans.

\begin{table}
\centering
\scalebox{0.80}{
\begin{tabular}{lllll}
\toprule
\multirow{2}{*}{\textbf{Model}} & \multicolumn{1}{r}{\multirow{2}{*}{\textbf{Size}}} & \multicolumn{3}{c}{\textbf{\passk} ($\%$)}                                                                        \\ 
\cmidrule(l){3-5}
                                & \multicolumn{1}{r}{}                                 & \multicolumn{1}{c}{\textit{k=$1$}} & \multicolumn{1}{c}{\textit{k=$10$}} & \multicolumn{1}{c}{\textit{k=$100$}}  \\ 
% \midrule
%\textbf{Model} & \textbf{Size} & \textbf{$\rm{pass}@1$} & \textbf{$\rm{pass}@10$} & \textbf{$\rm{pass}@100$}\\
\toprule
\multicolumn{5}{c}{\textit{\textbf{Model Size: \textasciitilde{}$100$M}}}                                                                                                                        \\
% TabNine                         & -                                                    & $2.58$                             & $4.35$                              & $7.59$                                \\
% GPT-CC                      & $125$M                                                 & $0.00$                             & $0.00$                              & $0.00$                               \\
% CodeT5                         & $60$M                                                 & $0.00$                             & $0.00$                              & $0.00$                                \\
% CodeT5                         & $220$M                                                 & $0.00$                             & $0.00$                              & $0.00$                                \\
GPT-Neo                         & $125$M                                                 & $0.75$                             & $1.88$                              & $2.97$                                \\
CodeParrot                      & $110$M                                                 & $3.80$                             & $6.57$                              & $12.78$                               \\
PyCodeGPT                       & $110$M                                                 & $\mathbf{8.33}$                             & $\mathbf{13.36}$                             & $19.13$                               \\
PolyCoder                       & $160$M                                                 & $2.13$                             & $3.35$                              & $4.88$                                \\
Codex                           & $12$M                                                  & $2.00$                             & $3.62$                              & $8.58$                                \\
Codex                           & $25$M                                                  & $3.21$                             & $7.1$                               & $12.89$                               \\
Codex                           & $42$M                                                  & $5.06$                             & $8.8$                               & $15.55$                               \\
Codex                           & $85$M                                                  & $8.22$                             & $12.81$                             & $\mathbf{22.40}$                                \\
AlphaCode(dec)                  & $13$M                                                  & $1.5$                              & $3.6$                               & $8.6$                                 \\
AlphaCode(dec)                  & $29$M                                                  & $3.4$                              & $5.8$                               & $11.2$                                \\
AlphaCode(dec)                  & $55$M                                                  & $4.2$                             & $8.2$                              & $16.9$                               \\
AlphaCode(dec)                  & $89$M                                                  & $4.3$                             & $12.2$                             & $20.0$                              \\ 
\hline
\multicolumn{5}{c}{\textit{\textbf{Model Size: \textasciitilde{}$500$M}}}                                                                                                                        \\
CodeT5$^{\mathcal{y}}$                         & $770$M                                                 & $12.09$                             & $19.24$                              & $30.93$                                \\
PolyCoder                       & $400$M                                                 & $2.96$                             & $5.29$                             & $11.59$                               \\
JuPyT5                          & $300$M                                                 & $5.40$                             & $15.46$                             & $25.60$                               \\
BLOOM                          & $560$M                                                 & $0.82$                             & $3.02$                             & $5.91$                               \\
Codex                           & $300$M                                                 & $13.17$                            & $20.37$                             & $36.27$                               \\
Codex                           & $679$M                                                 & $16.22$                            & $\mathbf{25.70}$                              & $\mathbf{40.95}$                               \\
AlphaCode(dec)                  & $302$M                                                 & $11.6$                             & $18.8$                              & $31.8$                                \\
AlphaCode(dec)                  & $685$M                                                 & $14.2$                             & $24.4$                              & $38.8$                                \\
CodeGen-Mono                    & $350$M                                                 & $12.76$                            & $23.11$                             & $35.19$                               \\
PanGu-Coder                     & $317$M                                                 & $\mathbf{17.07}$                            & $24.05$                             & $34.55$                               \\ 
\hline
\multicolumn{5}{c}{\textit{\textbf{Model Size: \textasciitilde{}$1$B}}}                                                                                                                          \\
% GPT-CC                      & $1.3$B                                                 & $0.00$                             & $0.00$                              & $0.00$                               \\
GPT-Neo                         & $1.3$B                                                 & $4.79$                             & $7.47$                              & $16.30$                               \\
CodeParrot                      & $1.5$B                                                 & $3.99$                             & $8.69$                              & $17.88$                               \\
BLOOM                      & $1.1$B                                                 & $2.48$                             & $5.93$                              & $9.62$                               \\
BLOOM                      & $1.7$B                                                 & $4.03$                             & $7.45$                              & $12.75$                               \\
InCoder$^{\mathcal{y}}$                      & $1.3$B                                                 & $11.09$                             & $16.14$                              & $24.20$                               \\
AlphaCode(dec)                   & $1.1$B                                                 & $17.1$                             & $28.2$                              & $45.3$                                \\
SantaCoder                      & $1.1$B                                                 & $\mathbf{18}$                               & $\mathbf{29}$                              & $\mathbf{49}$                                  \\ 
\hline
\multicolumn{5}{c}{\textit{\textbf{Model Size: \textasciitilde{}$5$B}}}                                                                                                                          \\
GPT-Neo                         & $2.7$B                                                 & $6.41$                             & $11.27$                             & $21.37$                               \\
PolyCoder                       & $2.7$B                                                 & $5.59$                             & $9.84$                              & $17.68$                               \\
Codex                           & $2.5$B                                                 & $21.36$                            & $35.42$                             & $59.50$                                \\
PanGu-Coder                     & $2.6$B                                                 & $23.78$                            & $35.36$                             & $51.24$                               \\ 
% \hline
% \multicolumn{5}{c}{\textit{\textbf{Model Size: \textasciitilde{}$6$B}}}                                                                                                                          \\
BLOOM                      & $3$B                                                 & $6.48$                             & $11.35$                              & $20.43$                               \\
BLOOM                      & $7.1$B                                                 & $7.73$                             & $17.38$                              & $29.47$                               \\
CodeGen-Mono                    & $2.7$B                                                 & $23.70$                            & $36.64$                             & $57.01$                               \\
CodeGen-Mono                    & $6.1$B                                                 & $\mathbf{26.13}$                            & $\mathbf{42.29}$                             & $\mathbf{65.82}$                               \\
GPT-J                           & $6$B                                                   & $11.62$                            & $15.74$                             & $27.74$                               \\
InCoder                         & $6.7$B                                                 & $15.2$                             & $27.8$                              & $47.0$                                \\ 
\hline
\multicolumn{5}{c}{\textit{\textbf{Model Size: \textgreater{}$10$B}}}                                                                                                                            \\

Codex                           & $12$B                                                  & $28.81$                            & $46.81$                             & $72.31$                               \\
CodeGen-Mono                    & $16.1$B                                                & $29.28$                            & $49.86$                             & $75.00$                               \\
GPT-NeoX                        & $20$B                                                  & $15.4$                             & $25.6$                              & $41.2$                                \\ 
% \hline
% \multicolumn{5}{c}{\textit{\textbf{Model Size: \textgreater{}$100$B}}}                                                                                                                           \\
LaMDA                           & $137$B                                                 & $14.0$                             & $-$                                 & $47.3$                                \\
BLOOM                           & $176$B                                                 & $15.52$                             & $32.20$                                 & $55.45$                                \\
PaLM-Coder                      & $540$B                                                 & $36.0$                             & $-$                                 & $88.4$                                \\
code-cushman-001                     & $-$                                                    & $33.5$                             & $54.3$                              & $77.4$                                \\
code-davinci-001                    & $-$                                                    & $39.0$                             & $60.6$                              & $84.1$                                \\
code-davinci-002                     & $-$                                                    & $\mathbf{47.0}$                             & $\mathbf{74.9}$                              & $\mathbf{92.1}$                                \\
\bottomrule
\end{tabular}
}
\caption{Performance on the HumanEval benchmark. $^{\mathcal{y}}$ denotes our reproduced results, while others are cited from the original papers. AlphaCode(dec) means the decoder-only version. We also compare the Codex models (code-cushman and code-davinci) provided by OpenAI API. We exclude the models that cannot pass any problem in the benchmark.}
\label{tab:models-performance-humaneval}
\end{table}

\section{What makes LLMs successful?}
\label{sec:analysis}
We have summarized the existing large language models for \nlcode. These LLMs vary in terms of architecture, size, and other characteristics, making it difficult to establish a completely fair comparison. We evaluate these LLMs on the HumanEval benchmark~\cite{codex} in a zero-shot manner to provide an intuitive comparison. HumanEval, proposed along with Codex, is one of the most popular benchmarks for the \nlcode task and consists of $164$ hand-written Python programming problems. Test cases are provided for each programming problem to evaluate the correctness of generated code. \passk is used as the evaluation metric\footnote{The details of \passk can be found in Appendix~\ref{apx:definition_passk}.}, which calculates the proportion of problems that can be correctly answered with $k$ tries. Table~\ref{tab:models-performance-humaneval} shows the results of different LLMs organized by the model size. Implementation details and the evaluation on the MBPP benchmark~\cite{mbpp} can be found in Appendix~\ref{apx:implementation_details}.

\begin{figure*}[t]
	\centering
	\begin{subfigure}[t]{0.48\linewidth}
		\centering
           \includegraphics[width=0.9\linewidth]{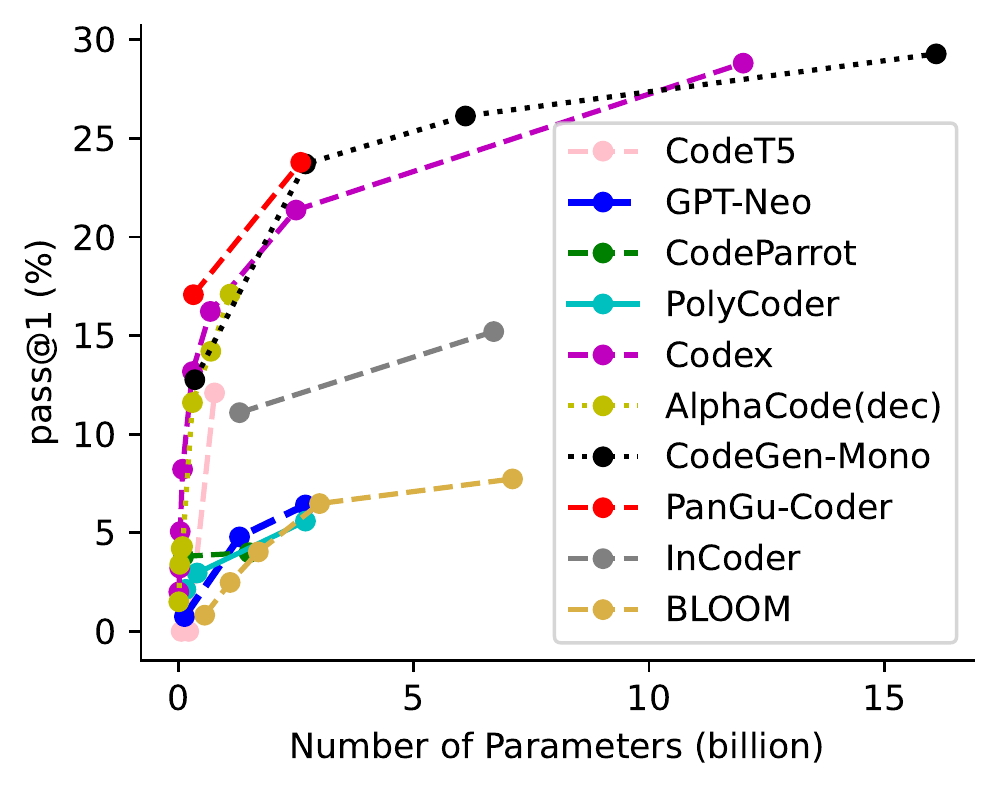}
		\caption{}
		\label{fig:params_performance_humaneval}
	\end{subfigure}
	\hfill
	\begin{subfigure}[t]{0.48\linewidth}
		\centering
		\includegraphics[width=0.9\linewidth]{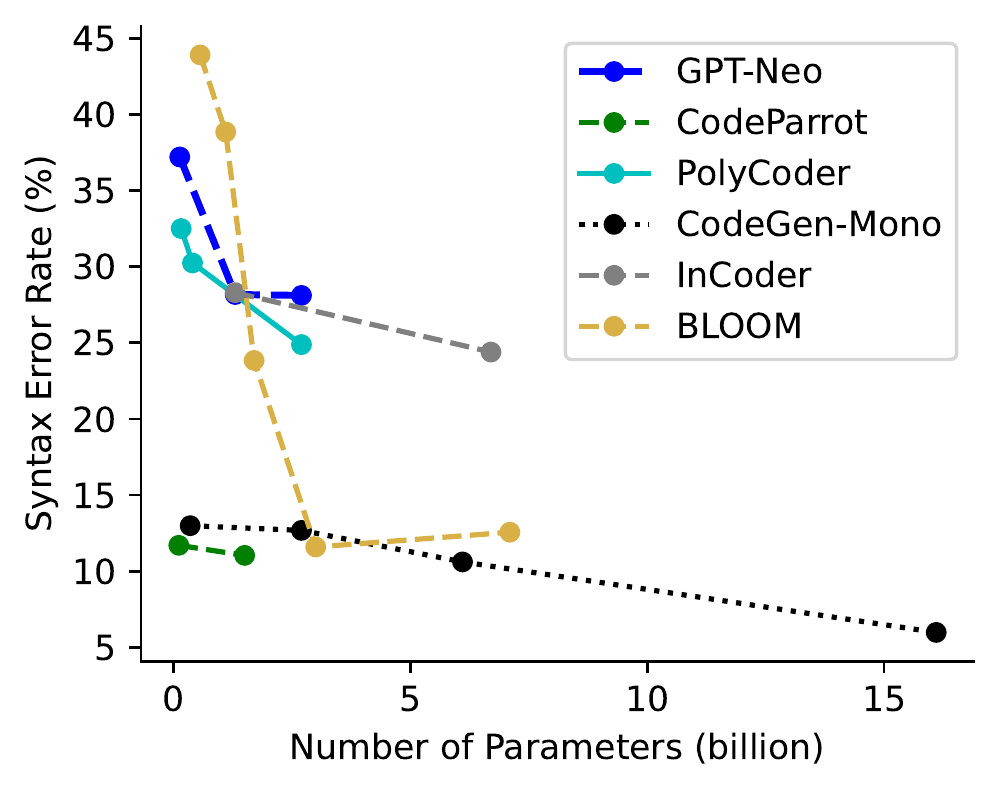}
		\caption{}
		\label{fig:syntax_error_params_humaneval}
	\end{subfigure}
	\caption{(a) ${\rm{pass}@}1$ and (b) syntax error rates on the HumanEval benchmark with various model sizes.}
	\label{fig:humaneval_results}
\end{figure*}

It can be observed from Table~\ref{tab:models-performance-humaneval} that the performance of existing LLMs varies widely on HumanEval, even for those with similar model sizes. Specifically, Codex~\cite{codex} holds the leading position in various model sizes, while a relatively small model, PyCodeGPT $110$M~\cite{cert}, achieves comparable results to Codex $85$M. Other larger models such as  AlphaCode~\cite{alphacode}, CodeGen-Mono~\cite{codegen}, and PanGu-Coder~\cite{pangu-coder} also exhibit impressive performance. Notably, InCoder~\cite{incoder} and SantaCoder~\cite{santacoder}, which use the FIM training method~\cite{fim}, also obtain remarkably decent results in the left-to-right generation setting.
The significant variation in performance leads us to the question: \emph{What makes LLMs successful in \nlcode?}
Given the diversity of these models in terms of design choices, we perform a thorough analysis and conclude the answer:
\textbf{\emph{Large Size, Premium Data, Expert Tuning}}. That is, large model and data size, high-quality data and expert hyper-parameter tuning are the key factors for the success of LLMs in the \nlcode task. In this section, we detail our observations and insights from the perspectives of model, data and tuning.

\subsection{Large Model Size}
As shown in Figure~\ref{fig:model-time-params} and Table~\ref{tab:models-performance-humaneval}, recent LLMs for \nlcode exhibit larger sizes and superior performance. This is consistent with prior findings that an increased number of model parameters can enhance model capabilities~\cite{gpt-2,lamda,palm}. We further demonstrate the correlation between model size and performance in Figure~\ref{fig:params_performance_humaneval}, which compares the ${\rm{pass}@}1$ results of $10$ representative models on the HumanEval benchmark. It is clear that larger models generally result in better performance. Furthermore, we also find that current models, regardless of size, still have the potential for improvement through further increases in size. Additional results on the HumanEval and MBPP benchmarks can be found in Appendix Figure~\ref{fig:params_performance_all}, which also support this conclusion. 

Additionally, we conduct an experiment on the HumanEval benchmark to examine the syntax error rates of the code generated by different models of varying sizes. Specifically, we make the models predict $10$ code samples for each programming problem, and then calculate the percentage of code samples that have syntax errors. As shown in Figure~\ref{fig:syntax_error_params_humaneval}, results indicate that larger models tend to have lower syntax error rates. It is noteworthy that the largest version of the CodeGen-Mono model exhibits a remarkably low rate of syntax errors, i.e., $6\%$. However, as evidenced by Figure~\ref{fig:params_performance_humaneval} and Table~\ref{tab:models-performance-humaneval}, the CodeGen-Mono model with $16$ billion parameters still has unsatisfactory performance in terms of \passk, e.g., ${\rm{pass}@}1$ to be $29\%$. This highlights the fact that the current limitation for large pre-trained models is the generation of semantically correct code.

\subsection{Large and Premium Data} 

As the sizes of LLMs increase in the field of \nlcode, the scale of the corpus used for training also increases. This highlights the importance of selecting and pre-processing high-quality data. In this section, we will discuss various commonly used data sources and pre-processing strategies that are essential for training LLMs.

Early models were trained using manually annotated data pairs of NL and code, and the data sources include CodeSearchNet~\cite{codesearchnet}, CoST~\cite{cost}, and XLCoST~\cite{xlcost}. 
However, manual annotation is labour-intensive and time-consuming. There are also models like GPT-3~\cite{gpt3}, GPT-Neo~\cite{gpt-neo}, and GPT-J~\cite{gpt-j} that are trained on the Pile~\cite{pile}, a large-scale unsupervised dataset.
However, these models have not yet demonstrated exceptional code generation capabilities due to the limited number of code files in the training corpus.
More recently, with the emergence of more powerful LLMs for \nlcode, larger-scale unlabelled code datasets have been proposed, including BigQuery~\cite{bigquery}, CodeParrot's corpus~\cite{codeparrot-data}, GitHub-Code~\cite{codeparrotdata-2}, and the Stack~\cite{thestack}, which are collected from general domain open-source websites like GitHub\footnote{\url{https://github.com}} and Stack Overflow\footnote{\url{https://stackoverflow.com}}. Furthermore, there are also specialized datasets proposed for different scenarios, for example, using Jupyter Notebooks or competition programming problems as a training corpus. Released datasets include Jupyter~\cite{codeparrotdata-3}, JuICe~\cite{juice}, APPS~\cite{apps}, and CodeNet~\cite{codenet}.

In order to ensure the quality of the training corpus, it is common for LLMs to perform data pre-processing on the significant amount of code in the collected data. We carefully review the data pre-processing methods of five powerful LLMs, including Codex~\cite{codex}, AlphaCode~\cite{alphacode}, CodeGen~\cite{codegen}, InCoder~\cite{incoder}, and PyCodeGPT~\cite{cert}, and identify several commonalities. One is the removal of likely auto-generated or unfinished code files, as they are deemed to be meaningless. Additionally, specific rules are employed to filter out uncommon code files. These rules include factors such as the repository star rating, the file size, the line length, and the alphanumeric rate. In summary, the goal of these pre-processing strategies is to achieve a code corpus that is unduplicated, complete, correct, clean, and general in nature.

\begin{figure}[t]
    \small
    \centering
    \includegraphics[width=0.92\linewidth]{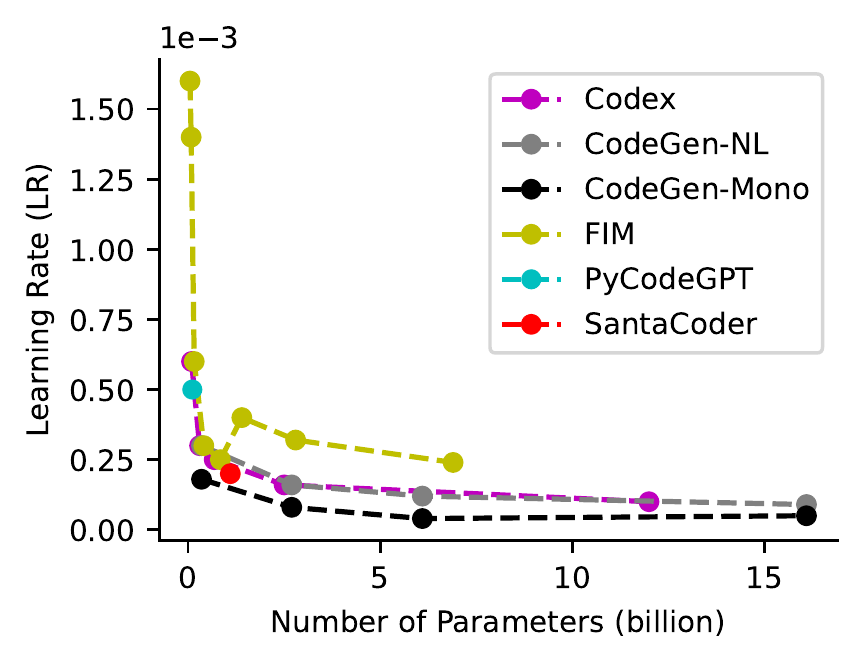}
    \caption{Learning rate of six advanced LLMs in terms of various model sizes.}
    \label{fig:params_learning_rate}
\end{figure}

\begin{figure}[t]
    \small
    \centering
    \includegraphics[width=0.9\linewidth]{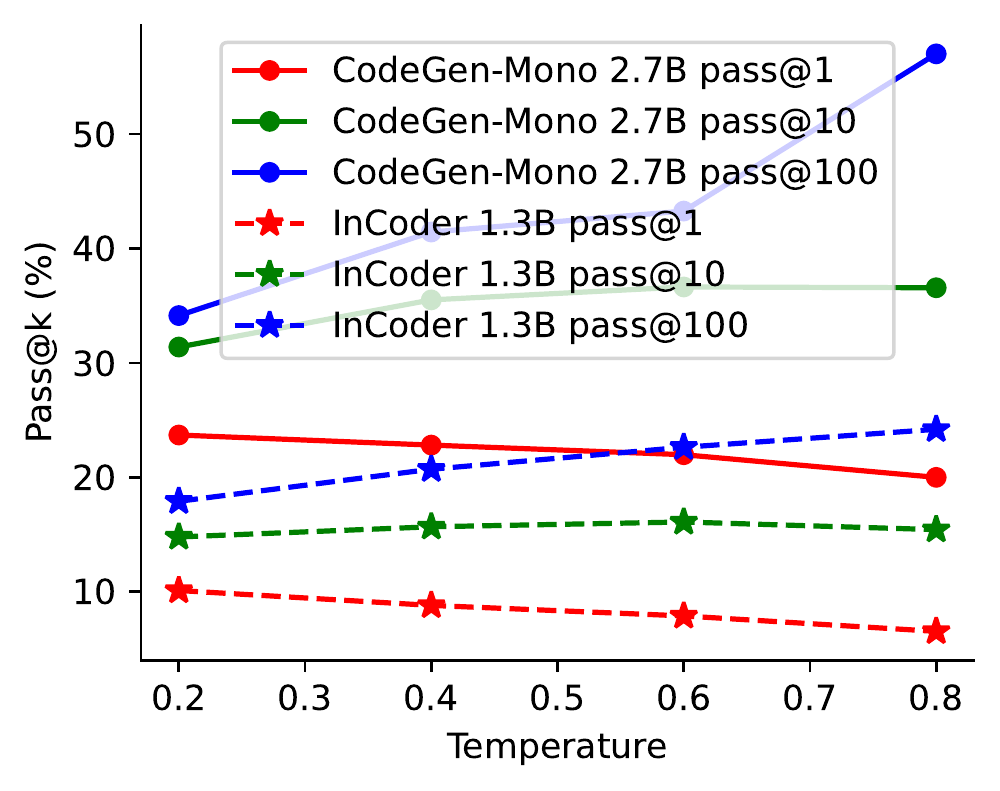}
    \caption{\passk on the HumanEval benchmark with different temperatures during model inference.}
    \label{fig:temperature_pass_rate}
\end{figure}

\begin{table*}
\centering
\scalebox{0.84}{
\rotatebox{0}{
\begin{tabular}{llllllllll} 
\toprule
\multirow{2}{*}{\textbf{Benchmark}} & \multirow{2}{*}{\textbf{Num.}} & \multirow{2}{*}{\textbf{P. NL}} & \multirow{2}{*}{\textbf{S. PL}} & \multicolumn{5}{c}{\textbf{Data Statistics}}                                                                                                                      & \multirow{2}{*}{\textbf{Scenario}} \\ 
\cline{5-9}
                                    &                                &                                 &                                 & \textit{T.N.} & \multicolumn{1}{c}{\textit{P.C.}} & \multicolumn{1}{c}{\textit{P.L.}} & \multicolumn{1}{c}{\textit{S.C.}} & \multicolumn{1}{c}{\textit{S.L.}} &                                    \\ 
\toprule
HumanEval~\citeyearpar{codex}                           & $164$                            & English                         & Python                          & $7.8$         & $450.6$                             & $13.7$                              & $180.9$                             & $6.8$                               & Code Exercise                       \\
MBPP~\citeyearpar{mbpp}                                & $974$                            & English                         & Python                          & $3.1$         & $78.6$                              & $1.0$                               & $181.1$                             & $6.7$                               & Code Exercise                       \\
APPS~\citeyearpar{apps}                                & $5,000$                          & English                         & Python                          & $21.0$        & $1743.4$                            & $41.6$                              & $473.8$                             & $21.4$                              & Competitions                       \\
CodeContests~\citeyearpar{alphacode}                        & $165$                            & English                         & Multi.                          & $203.7$       & $1989.2$                            & $66.4$                              & $2239.3$                            & $92.1$                              & Competitions                       \\
DS-1000~\citeyearpar{ds-1000}                             & $1,000$                          & English                         & Python                          & $1.6$         & $879.1$                             & $31.6$                              & $137.4$                             & $5.0$                               & Data Science                       \\
DSP~\citeyearpar{dsp}                                 & $1,119$                          & English                         & Python                          & $2.1$         & $756.9$                             & $17.8$                              & $226.3$                             & $7.6$                               & Data Science                       \\
MBXP~\citeyearpar{mbxp}                                & $974^*$                         & English                         & Multi.                          & $3.1$         & $419.9$                             & $14.8$                              & $-$                                 & $-$                                 & Multilingual                       \\
MBXP-HumanEval~\citeyearpar{mbxp}                      & $164^*$                         & English                         & Multi.                          & $7.8$         & $825.6$                             & $30.0$                              & $-$                                 & $-$                                 & Multilingual                       \\
HumanEval-X~\citeyearpar{codegeex}                         & $164^*$                         & English                         & Multi.                          & $7.8$         & $468.4$                             & $15.5$                              & $264.6$                             & $12.1$                              & Multilingual                       \\
MultiPL-HumanEval~\citeyearpar{multipl-e}                   & $164^*$                         & English                         & Multi.                          & $7.8$         & $453.9$                             & $13.0$                              & $-$                                 & $-$                                 & Multilingual                       \\
MultiPL-MBPP~\citeyearpar{multipl-e}                        & $974^*$                         & English                         & Multi.                          & $3.1$         & $181.2$                             & $5.4$                               & $-$                                 & $-$                                 & Multilingual                       \\
PandasEval~\citeyearpar{cert}                          & $101$                            & English                         & Python                          & $6.5$         & $244.5$                             & $7.2$                               & $46.2$                              & $1.3$                               & Public Library                     \\
NumpyEval~\citeyearpar{cert}                           & $101$                            & English                         & Python                          & $3.5$         & $222.9$                             & $7.0$                               & $29.9$                              & $1.1$                               & Public Library                     \\
TorchDataEval~\citeyearpar{apicoder}                       & $50$                             & English                         & Python                          & $1.1$         & $329.0$                             & $8.6$                               & $50.7$                              & $1.3$                               & Private Library                    \\
MTPB~\citeyearpar{codegen}                                & $115$                            & English                         & Python                          & $-$           & $72.7$                              & $1.0$                               & $-$                                 & $-$                                 & Multi-Turn                         \\
ODEX~\citeyearpar{odex}                                & $945$                            & Multi.                          & Python                          & $1.8$         & $26.6$                              & $2.0$                               & $50.4$                              & $1.9$                               & Open-Domain                        \\
BIG-Bench~\citeyearpar{big-bench}                           & $32$                             & English                         & Python                          & $4.7$         & $341.8$                             & $3.0$                               & $-$                                 & $-$                                 & Code Exercise                       \\
\bottomrule
\end{tabular}
}
}
\caption{Summary of $17$ benchmarks for \nlcode. Num. denotes the number of instances in the benchmark, P.NL denotes \underline{P}roblem description's \underline{N}atural \underline{L}anguage, S.PL denotes code \underline{S}olution's \underline{P}rogramming \underline{L}anguage, and T.N. denotes the average \underline{N}umber of \underline{T}est cases. P.C. and P.L. (S.C. and S.L.) stand for the average number of \underline{C}haracters and \underline{L}ines in \underline{P}roblem description (code \underline{S}olution). $^*$ denotes the number of instances per programming language.}
\label{tab:benchmarks}
\end{table*}

\subsection{Expert Tuning} \label{sec:masterful_tuning}

Training an excellent model requires careful consideration of various design choices and hyper-parameters. After reviewing the existing $27$ LLMs (summary in Appendix Table~\ref{tab:models-details}), we have the following findings. Firstly, these LLMs share some common settings. For example, we observe that the optimizer of the current models is almost all Adam~\cite{adam} or its variants~\cite{adamw}.
We also find that initializing with other natural language models yields no noticeable gain compared to training from scratch, except for accelerating convergence~\cite{codex}. Furthermore, there are several hyper-parameters that require expert tuning, such as learning rate, batch size, window size, warmup steps, gradient accumulation steps, and sampling temperature.
For the learning rate, we analyze its correlation with model size using six powerful LLMs, as shown in Figure~\ref{fig:params_learning_rate}. We observe that the learning rate becomes smaller as the model gets larger.
To explore the effects of temperature, in Figure~\ref{fig:temperature_pass_rate}, we report the performance of two models using multiple temperatures on HumanEval. One observation is that higher temperature leads to lower ${\rm{pass}@}1$ and higher ${\rm{pass}@}100$, which suggests that a higher temperature makes LLMs generate more diverse predictions and vice versa.
Besides, some studies~\cite{bigtransformers} have shown that window size is a key factor. An interesting finding is that the small model with a large window size sometimes outperforms the large model with a small window size (details in Appendix~\ref{apx:context_window_vs_performance}).
In addition, powerful LLMs usually train a new tokenizer on code corpus primarily using two techniques: Byte-level Byte-Pair-Encoding~\cite{gpt-2} and SentencePiece~\cite{sentencepiece}. A new tokenizer can be more effective and accurate in splitting code content into tokens. 
These proven tuning techniques will serve as valuable references for training more powerful LLMs.

\section{Benchmarks and Metrics} \label{sec:benchmarks}

To evaluate the \nlcode task, high-quality benchmarks and reliable metrics are fundamental and essential. In this section, we provide a brief overview of current benchmarks and metrics, as well as our observations and the open challenges.

We summarize $17$ well-studied \nlcode benchmarks in Table~\ref{tab:benchmarks}, where we can find that each of these benchmarks has its own characteristics regarding size, language, complexity, and scenario. We observe that most benchmarks contain a limited number of instances. For example, the widely used HumanEval and MBPP have $164$ and $974$ instances, respectively. This is because these benchmarks are typically hand-written to ensure that LLMs have not seen them during training. In the era of large language models, it is crucial to avoid data leakage when creating new benchmarks. 
Additionally, most current benchmarks have their problem descriptions in English and code solutions in Python. Recently, several multi-lingual benchmarks have been proposed, such as MBXP~\cite{mbxp}, HumanEvalX~\cite{codegeex}, and MultiPL~\cite{multipl-e}, which cover multiple programming languages, and ODEX~\cite{odex}, which covers multiple natural languages. Details of multi-lingual benchmarks are listed in Appendix Table~\ref{tab:benchmarks-multilingual}. 
Furthermore, benchmarks have been proposed for other practical scenarios, such as data science~\cite{ds-1000}, public library~\cite{cert}, private library~\cite{apicoder}, multi-turn program synthesis~\cite{codegen}, and code security~\cite{securityeval}. 
For execution-based benchmarks, comprehensive test cases with complete coverage of the generated program can ensure the trustworthiness of evaluation results.
As a reference, the average number of test cases for each benchmark, as well as the length statistics of the problem descriptions and solutions are also provided in Table~\ref{tab:benchmarks}. 

Manually evaluating the generated code is impractical, which calls for the need for automatic metrics. The above mentioned benchmarks all provide test cases for execution-based evaluation, where metrics such as \passk~\cite{codex}, $n@k$~\cite{alphacode}, test case average~\cite{apps}, and execution accuracy~\cite{incontext-1} can be used. However, this approach has stringent requirements for the quality of test cases and can only evaluate executable code. For non-executable code, metrics like BLEU~\cite{bleu}, ROUGE~\cite{rouge}, and CodeBLEU~\cite{codebleu} are used, while they can not precisely evaluate the correctness of the code. So far, there are many open challenges in designing metrics to evaluate various aspects of code, such as vulnerability, maintainability, clarity, execution complexity, and stability.

\section{Challenges and Opportunities} \label{sec:challenges-and-future-works}
Our investigations have revealed that advances in LLMs for \nlcode have a considerable impact on both academia and industry. Despite this progress, there are still numerous challenges that need to be addressed, offering ample opportunities for further research and applications. In this section, we explore the challenges and opportunities in terms of the ability gap between LLMs and humans.

\paragraph{Understanding Ability}

The inherent flexibility of natural language allows for a variety of expressions to convey functional requirements. Humans are able to understand various descriptions at different levels of abstraction. In contrast, current LLMs tend to be sensitive to the given context, which may cause unexpected performance degradation~\cite{sensitive}. In addition, LLMs may struggle when faced with complex problems that have numerous conditions and requirements~\cite{grounded-copilot,copilot-5}. We believe exploring the understanding abilities of LLMs is a crucial research direction. One potential solution is to break down complex problems into multiple steps, as is commonly done in reasoning tasks~\cite{reasoning-2}.

\paragraph{Judgement Ability}

Humans have the ability to determine whether they can solve a programming problem or not. While current models will always return a solution even if there is no answer to the problem, due to the fact that they are trained by unsupervised causal language modeling objective. This can cause problems in practical applications. To improve the judgment ability of LLMs, researchers have employed reinforcement learning to leverage user feedback, as seen in models like InstructGPT~\cite{instructgpt} and ChatGPT\footnote{\url{https://chat.openai.com}}. However, collecting high-quality feedback for code is costly and challenging. There are also ongoing studies~\cite{codet,ispeakuverify} exploring the possibility of self-validation for LLMs, which is also a promising research direction.

\paragraph{Explanation Ability}

It is widely acknowledged that human developers possess the ability to interpret the meaning of the code they write, which is crucial for educational purposes and software maintenance. Recent studies showed that LLMs have the potential to automatically generate code explanations. \citet{explanation-1} proposed using LLMs to generate code explanations for students during their learning process, and \citet{explanation-2} proposed explaining numerous aspects of a given code snippet using Copilot. Further research and explorations are necessary to fully realize the potential of LLMs in this regard.

\paragraph{Adaptive Learning Ability}

A fundamental difference between current large language models and humans is their ability to adapt to new and updated knowledge. Human developers possess a unique ability to quickly search and learn new materials, such as programming documentation, and adapt to changes in APIs with relative ease. However, re-training or fine-tuning LLMs requires significant effort and resources. This issue has inspired a number of recent studies, such as DocCoder~\cite{retri-2} and APICoder~\cite{apicoder}, which utilize retrieval-based methods to provide extra or updated knowledge during model inference. Despite these advancements, it remains an open challenge to endow LLMs with the powerful learning capabilities humans possess.

\paragraph{Multi-tasking Ability}

Large language models have been applied to a variety of code-related tasks, such as code repair~\cite{repair-1,repair-2}, code search~\cite{code-retrieval}, and code review~\cite{code-review} as well as non-code tasks that can be formatted in a code-like manner, such as mathematics~\cite{math-2,math-1} and chemistry~\cite{chem-1,edu-5-chem-2}. However, there are differences between LLMs and human abilities in terms of multi-tasking. Humans can seamlessly switch between tasks, while LLMs may require sophisticated prompt engineering~\cite{prompt-engineering}. Another evidence is that LLMs lack the ability to quickly master multiple programming languages~\cite{codegeex} as humans do. These limitations highlight areas for future research.

\section{Conclusion} \label{sec:conclusion}

In this paper, we survey $27$ existing large language models for \nlcode, and draw a thorough analysis of the underlying reasons for their success.
We also provide a detailed review of benchmarks and metrics.
Regarding the gap between models and humans, we present ongoing challenges and opportunities.
In addition, we have developed a website to track the latest findings in this field.
We hope this survey can contribute to a comprehensive overview of the field and promote its thriving evolution.

\section*{Limitations} \label{sec:limitations}
In this paper, we thoroughly investigate the existing large language models for \nlcode, and summarize them from diverse perspectives with our own thinking. However, as this field is evolving so rapidly, there may be aspects that we have overlooked, or some new works that we have not covered. To mitigate this issue, we have created a website to track the latest progress through crowd-sourcing, hoping that it will continually contribute to the development of the field.
Besides, the existing LLMs possess their own characteristics in terms of model size, architecture, corpus, pre-processing, tokenizer, hyper-parameters, and training platforms. Also, some of them are currently not publicly available, such as AlphaCode~\cite{alphacode} and PaLM-Coder~\cite{palm}. Therefore, it is almost impractical to conduct a completely fair comparison. We tried our best to show a kind of comparison on the popular HumanEval and MBPP benchmarks, hoping that it can provide clues to the differences in performance of different LLMs. In addition, evaluating LLMs has a high cost in computational resources. 
We thus have made all files generated by the LLMs publicly available on \url{https://nl2code.github.io}.

% ---
% Entries for the entire Anthology, followed by custom entries
% \normalem
\bibliography{anthology,custom}
\bibliographystyle{acl_natbib}

\newpage
\appendix
\clearpage

\section{Related Surveys} \label{apx:related-survey}
Previous surveys on the topic of code intelligence~\cite{survey-1,survey-3,survey_ci_11,survey-14} and code generation~\cite{survey-9,survey-7,survey-2} have primarily focused on early methodologies such as the use of programming templates~\cite{survey-6,survey_12}, neural models based on CNN, RNN, and LSTM architectures~\cite{survey-1,survey-13}, and small-scale Transformer models that require labelled data for training~\cite{transformer-1,transformer-2}. However, with the advancement of model size, Transformer-based models have demonstrated exceptional performance in \nlcode tasks and have given rise to the development of more capable code generation models. In light of this, there exists a clear need for a comprehensive survey of large language models for \nlcode tasks to bridge this gap in knowledge. This study endeavours to fulfill this need by providing a thorough analysis of the successful LLMs and a detailed review of \nlcode benchmarks and metrics.
We also present the ongoing challenges and opportunities regarding the ability gap between LLMs and humans.

Finally, we would like to highlight some criteria for our survey. First, we only refer to official papers to investigate the size of the models. For example, Codex reported the model with a maximum size of $12$B in the paper, but later trained larger ones. In this case, we only consider the $12$B model as the largest one. In addition, the publication dates of the models in Figure~\ref{fig:model-time-params} are taken from official papers or blogs.

\section{An Online Website} \label{apx:website}
To keep tracking the latest progress of LLMs for \nlcode, we have developed an online real-time update website at \url{https://nl2code.github.io}. We have collected as many of the latest research works as possible on this website. Everyone is allowed to contribute to the website by pulling requests on GitHub. 
This website also includes features such as fuzzy search and custom tag categories, which will facilitate researchers to find the papers they want quickly. We hope this website can assist researchers and developers in related fields and contribute to its advancement.

\begin{table}[t]
\centering
\scalebox{0.88}{
\begin{tabular}{lllll} 
\toprule
\multirow{2}{*}{\textbf{Model}} & \multicolumn{1}{r}{\multirow{2}{*}{\textbf{Size}}} & \multicolumn{3}{c}{\textbf{\passk}}                                                                        \\ 
\cmidrule(l){3-5}
                                & \multicolumn{1}{r}{}                                 & \multicolumn{1}{c}{\textit{k=$1$}} & \multicolumn{1}{c}{\textit{k=$10$}} & \multicolumn{1}{c}{\textit{k=$100$}}  \\ 
\toprule
\multicolumn{5}{c}{\textit{\textbf{Model Size: \textasciitilde{}$100$M}}}                                                                                                                        \\
% GPT-CC                          & $125$M                                                 & $0.00$                                & $0.00$                                 & $0.00$                                   \\
% CodeT5                          & $60$M                                                  & $0.00$                             & $0.00$                              & $0.00$                                \\
% CodeT5                          & $220$M                                                 & $0.00$                             & $0.00$                              & $0.00$                                \\
GPT-Neo$^\mathcal{y}$                         & $125$M                                                 & $0.26$                             & $2.15$                              & $7.96$                                \\
% CodeGPT                         & $124$M                                                 & $0.00$                             & $0.00$                              & $0.00$                                \\

CodeParrot$^\mathcal{y}$                      & $110$M                                                 & $0.48$                             & $3.89$                              & $15.93$                               \\
PyCodeGPT$^\mathcal{y}$                       & $110$M                                                 & $\mathbf{9.39}$                             & $\mathbf{28.37}$                             & $\mathbf{48.71}$                               \\
PolyCoder$^\mathcal{y}$                       & $160$M                                                 & $1.08$                             & $6.67$                              & $18.97$                               \\ 
\hline
\multicolumn{5}{c}{\textit{\textbf{Model Size: \textasciitilde{}$500$M}}}                                                                                                                        \\
CodeT5$^\mathcal{y}$                          & $770$M                                                 & $\mathbf{15.78}$                             & $38.63$                              & $50.35$                                \\
PolyCoder$^\mathcal{y}$                       & $400$M                                                 & $1.31$                             & $7.98$                              & $21.55$                               \\
BLOOM$^\mathcal{y}$                       & $560$M                                                 & $0.26$                             & $2.04$                              & $8.90$                               \\
CodeGen-Mono$^\mathcal{y}$                    & $350$M                                                 & $15.44$                             & $\mathbf{42.50}$                             & $\mathbf{64.40}$                               \\
\hline
\multicolumn{5}{c}{\textit{\textbf{Model Size: \textasciitilde{}$1$B}}}                                                                                                                          \\
% GPT-CC                          & $1.3$B                                                 & $0.00$                                & $0.00$                                 & $0.00$                                   \\
GPT-Neo$^\mathcal{y}$                         & $1.3$B                                                 & $3.77$                             & $16.26$                              & $29.51$                                \\
CodeParrot$^\mathcal{y}$                      & $1.5$B                                                 & $1.29$                             & $8.66$                              & $27.17$                               \\ 
BLOOM$^\mathcal{y}$                      & $1.1$B                                                 & $1.90$                             & $9.20$                             & $23.42$                               \\
BLOOM$^\mathcal{y}$                      & $1.7$B                                                 & $3.16$                             & $14.23$                             & $31.38$                               \\
InCoder$^\mathcal{y}$                & $1.3$B                                                 & $\mathbf{10.00}$                            & $\mathbf{34.02}$                             & $\mathbf{55.50}$                               \\
SantaCoder$^\mathcal{y}$                      & $1.1$B                                                 & $3.65$                             & $21.33$                             & $41.92$                               \\
\hline
\multicolumn{5}{c}{\textit{\textbf{\textbf{Model Size: \textasciitilde{}$5$B}}}}                                                                                                                 \\
GPT-Neo$^\mathcal{y}$                         & $2.7$B                                                 & $5.89$                             & $23.09$                              & $44.26$                                \\
PolyCoder$^\mathcal{y}$                       & $2.7$B                                                 & $4.39$                             & $17.99$                             & $38.17$                               \\
BLOOM$^\mathcal{y}$                      & $3$B                                                 & $2.25$                             & $13.58$                             & $32.08$                               \\
BLOOM$^\mathcal{y}$                      & $7.1$B                                                 & $1.01$                             & $7.91$                             & $24.12$                               \\
CodeGen-Mono$^\mathcal{y}$                    & $2.7$B                                                 & $28.80$                            & $60.73$                             & $\mathbf{75.41}$                               \\
CodeGen-Mono$^\mathcal{y}$                    & $6.1$B                                                 & $\mathbf{33.70}$                                & $\mathbf{62.70}$                                 & $70.25$                                   \\
% \hline
% \multicolumn{5}{c}{\textit{\textbf{Model Size: \textasciitilde{}$6$B}}}                                                                                                                          \\
GPT-J$^\mathcal{y}$                           & $6$B                                                   & $11.30$                                & $35.62$                                 & $53.63$                                   \\
InCoder                         & $6.7$B                                                   & $21.3$                             & $46.5$                              & $66.2$                                \\ 
\hline
\multicolumn{5}{c}{\textit{\textbf{Model Size: \textgreater{}$10$B}}}                                                                                                                            \\
CodeGen-Mono                    & $16.1$B                                                & $42.4$                             & $65.8$                              & $79.1$                                \\ 
% \hline
% \multicolumn{5}{c}{\textit{\textbf{\textbf{Model Size: \textgreater{}$100$B}}}}                                                                                                                  \\
cushman-001                     & $-$                                                    & $45.9$                             & $66.9$                              & $79.9$                                \\
davinci-001                     & $-$                                                    & $51.8$                             & $72.8$                              & $84.1$                                \\
davinci-002                     & $-$                                                    & $\mathbf{58.1}$                             & $\mathbf{76.7}$                              & $\mathbf{84.5}$                                \\
\bottomrule
\end{tabular}
}
\caption{The performance of LLMs on the MBPP benchmark. $^\mathcal{y}$ denotes our reproduced results, while others are taken from \citet{codet}. We omit CodeGPT, GPT-CC, and PLBART as their numbers are zero.}
\label{tab:models-performance-mbpp}
\end{table}

\section{Experimental Setup} \label{apx:experiment_setup}

In this section, we will first present the definition of \passk, followed by the details of the experiments conducted on two benchmarks, namely HumanEval~\cite{codex} (results in Table~\ref{tab:models-performance-humaneval}) and MBPP~\cite{mbpp} (results in Table~\ref{tab:models-performance-mbpp}).

\subsection{Definition of \passk} \label{apx:definition_passk}
We use \passk as our metric for evaluation. For each programming problem, we sample $n$ candidate code solutions and then randomly pick ${k}$ of them. If any of the ${k}$ code solutions pass the given test cases, the problem can be regarded as solved. So \passk is the proportion of solved problems in the benchmark~\cite{codex}.
Formally, assuming that the number of correct ones in ${k}$ samples is $c$, \passk$=1$ if $n-c<{k}$; otherwise, \passk$=1-\prod\nolimits_{i=n-c+1}^{n} (1-{k}/i)$.
We chose \passk as our primary evaluation metric because it offers a completely precise evaluation of code accuracy by executing test cases, while other metrics mentioned in Section~\ref{sec:benchmarks} either originate from \passk or have lower precision.

\subsection{Implementation Details} \label{apx:implementation_details}

For HumanEval, we use the original benchmark\footnote{\url{https://github.com/openai/human-eval/blob/master/data/HumanEval.jsonl.gz}}. 
Most results in Table~\ref{tab:models-performance-humaneval} are taken from the original papers, while we reproduce the results of GPT-CC, PLBART, CodeT5, and InCoder $1.3$B by strictly following the same experimental setup as the other models.
In detail, we set the sample number to $200$, the maximum length of newly generated tokens to $200$, and \verb|top_p| to $0.95$.
We set the temperature from $0.1$ to $1.0$ with an interval of $0.1$, and report the best performance across these temperatures.

For MBPP, we use the version from \citet{codet}\footnote{\url{https://github.com/microsoft/CodeT/blob/main/CodeT/data/dataset/mbpp_sanitized_for_code_generation.jsonl}}. In Table~\ref{tab:models-performance-mbpp}, the results of InCoder $6.7$B and models larger than $10$B are taken from \citet{codet}, while we reproduced other results. Specifically, we set the sample number to $100$, the maximum length of newly generated tokens to $200$, \verb|top_p| to $0.95$, and the temperature to $0.8$.

For the two benchmarks above, we employ the same post-processing strategy. Following Codex~\cite{codex}, we terminate the sampling process when one of the following sequences is encountered in the generated code: `\verb|\nclass|', `\verb|\ndef|', `\verb|\n#|', `\verb|\n@|', `\verb|\nif|', and `\verb|\nprint|'. In our experiments, CodeT5 $770$M refers to the version\footnote{\url{https://huggingface.co/Salesforce/codet5-large-ntp-py}} with the causal language modeling objective. For good reproducibility and further research, we have made our code and the generated results of the LLMs on HumanEval and MBPP publicly available on our website.

\section{Context Window vs. Performance} \label{apx:context_window_vs_performance}

\begin{figure}[t]
    \small
    \centering
    \includegraphics[width=0.9\linewidth]{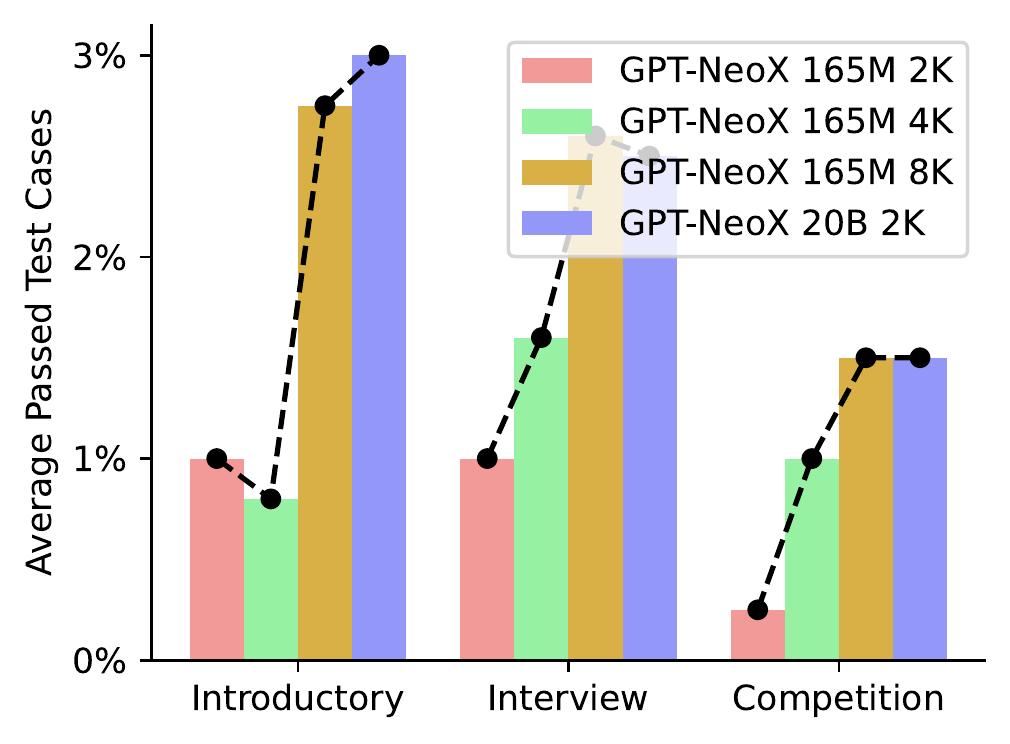}
    \caption{Performance of GPT-NeoX with different model sizes ($165$M and $20$B) and context windows ($2$K, $4$K, and $8$K) on the APPS benchmark.}
    \label{fig:context_window_performance}
\end{figure}

Recent work \cite{bigtransformers} claimed that the size of the context window plays a vital role in enhancing the performance of LLMs for \nlcode. Specifically, experiments are conducted on the APPS benchmark~\cite{apps} with GPT-NeoX~\cite{gpt-neox-20b}, and we visualize the results in Figure~\ref{fig:context_window_performance}. It is found that the $165$M version model with an $8,000$ context window is comparable to the $20$B version model with a $2,000$ context window. This observation illustrates that the context window also needs to be considered when training the model.

\begin{table*}[t]
\centering
\scalebox{0.82}{
\rotatebox{0}{
\begin{tabular}{llll} 
\toprule
\textbf{Products} & \textbf{Model} & \multicolumn{1}{c}{\textbf{Supported PLs}}                                                                                                                                                                                     & \textbf{Supported IDEs}                                                                                                                                                                                                                                \\ 
\toprule
tabnine~\citeyearpar{tabnine}           & $-$              & \begin{tabular}[c]{@{}l@{}}Python, Java, Javascript, TypeScript,\\Go, Ruby, PHP, C\#, C, C++,~Swift,~\\Perl,~Rust,~CSS,~Angular, Dart, React,\\Haskell, HTML, Kotlin, Matlab, Sass,\\NodeJS, Objective C, Scala,~\end{tabular} & \begin{tabular}[c]{@{}l@{}}VS Code, Visual Studio, IntelliJ IDE,\\Neovim, Sublime, PyCharm,~Rider,\\WebStorm, Android Studio,~Emacs,\\Vim, PhpStorm, RubyMine,~DataGrip,\\Jupyter Notebook,~JupyterLab, Clion,\\AppCode, Eclipse,~GoLand\end{tabular}  \\ 
\hline
aiXcoder~\citeyearpar{aiXcoder}          & $-$              & \begin{tabular}[c]{@{}l@{}}Python, Java, JavaScript, Typescript, \\Go, PHP, C, C++\end{tabular}                                                                                                                                & \begin{tabular}[c]{@{}l@{}}VS Code, IntelliJ IDEA, PyCharm,~\\STS3,~WebStorm,~Rider, Clion, STS4\\Android Studio,~PhpStorm, Eclipse,\\GoLand\end{tabular}                                                                                              \\ 
\hline
IntelliCode~\citeyearpar{intellicode}       & $-$              & \begin{tabular}[c]{@{}l@{}}Python, Java, JavaScript, TypeScript,\\C\#, C++, SQL Server, XAML\end{tabular}                                                                                                                      & VS Code, Visual Studio                                                                                                                                                                                                                                 \\ 
\hline
Diffblue Cover~\citeyearpar{diffblue-cover}    & $-$              & Java                                                                                                                                                                                                                           & IntelliJ IDEA, CLI Tool                                                                                                                                                                                                                                \\ 
\hline
Copilot~\citeyearpar{copilot}           & Codex          & \begin{tabular}[c]{@{}l@{}}Python, Java, JavaScript, TypeScript,\\Go, Ruby, Julia,~PHP,~C\#,~C++,~Swift,\\Perl,~PowerShell,~R, Rust,~CSS,~SQL,~\\JSON, HTML, SCSS, Less,~.NET,\\Markdown,~T-SQL\end{tabular}                   & \begin{tabular}[c]{@{}l@{}}VS Code, Visual Studio,~Neovim,\\JetBrains IDE\end{tabular}                                                                                                                                                                 \\ 
\hline
Cosy~\citeyearpar{cosy}              & $-$              & Java                                                                                                                                                                                                                           & IntelliJ IDEA                                                                                                                                                                                                                                          \\ 
\hline
CodeWhisperer~\citeyearpar{codewhisperer} & $-$              & \begin{tabular}[c]{@{}l@{}}Python, Java, JavaScript, TypeScript,\\C\#\end{tabular}                                                                                                                                             & \begin{tabular}[c]{@{}l@{}}VS Code, JetBrains IDE, AWS Cloud9,\\AWS Lambda\end{tabular}                                                                                                                                                                \\ 
\hline
CodeGenX~\citeyearpar{codegenx}          & GPT-J          & Python                                                                                                                                                                                                                         & VS Code                                                                                                                                                                                                                                                \\ 
\hline
CodeGeeX~\citeyearpar{codegeex}          & CodeGeeX       & \begin{tabular}[c]{@{}l@{}}Python, Java, JavaScript, TypeScript,\\Go,~PHP,~C\#,~C,~C++,~Perl,~Rust, CSS,\\SQL, HTML,~Kotlin, Shell,~R, Cuda, \\Objective C,~Objective C++, Pascal,\\Tex,~Fortran, Lean,~Scala\end{tabular}     & \begin{tabular}[c]{@{}l@{}}VS Code, IntelliJ IDEA, PyCharm,\\WebStorm,~Android Studio,~Rider,\\RubyMine, Clion, AppCode,~Aqua,\\DataGrip,~GoLand, DataSpell\end{tabular}                                                                               \\ 
\hline
FauPilot~\citeyearpar{fauxpilot}          & CodeGen        & Python, Java, Javascript, Go, C, C++                                                                                                                                                                                           & $-$                                                                                                                                                                                                                                                      \\
\bottomrule
\end{tabular}
}
}
\caption{Summary of products powered by LLMs. PLs and IDEs refer to programming languages and integrated development environments, respectively. The information for these products was recorded on December $27$, $2022$.}
\label{tab:products}
\end{table*}

\begin{figure*}[t]
    \small
    \centering
    \includegraphics[width=1.0\linewidth]{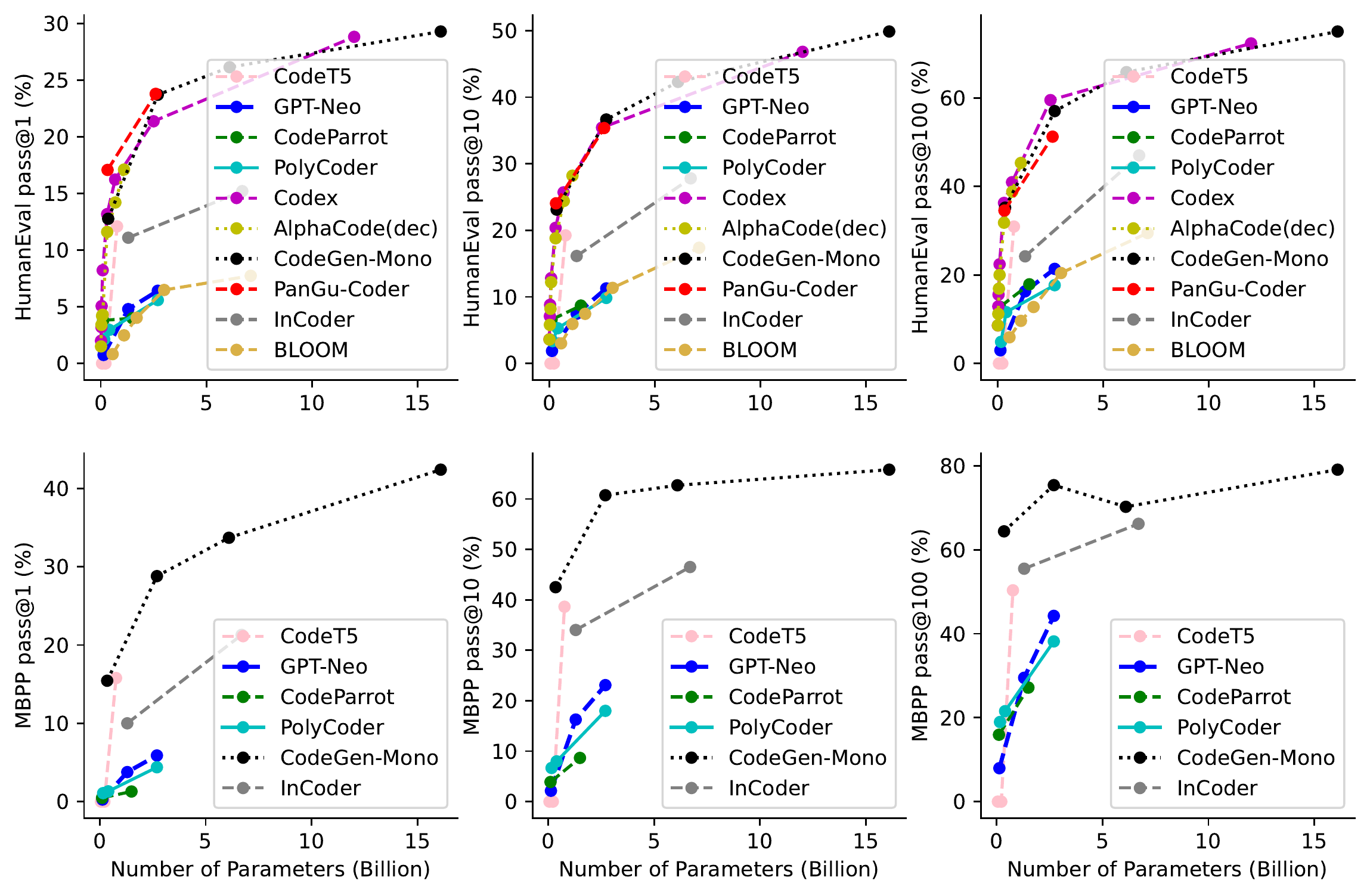}
    \caption{Performance of LLMs with varying parameter sizes on the HumanEval and MBPP benchmarks.}
    \label{fig:params_performance_all}
\end{figure*}

\begin{table*}[t]
\centering
\scalebox{0.92}{
\rotatebox[origin=c]{-90}{
\begin{tabular}{|l|l|l|l|l|l|l|l|l|l|l|l|l|l|l|l|} 
\hline\hline
\rowcolor[rgb]{0.894,0.894,0.894} {\cellcolor[rgb]{0.894,0.894,0.894}}                                          & \multicolumn{2}{c|}{\textbf{\textit{Data}}}       & \multicolumn{11}{c|}{\textbf{\textit{Model Hyper-parameters }}}                                                                                                                                                                                                        & \multicolumn{2}{c}{\textbf{\textit{Training}}}  \\ 
\rowcolor[rgb]{0.894,0.894,0.894} \multirow{-2}{*}{{\cellcolor[rgb]{0.894,0.894,0.894}}\textbf{\textit{Model}}} & \textbf{\textit{de.}} & \textbf{\textit{token.}} & \textbf{\textit{opti.}} & \textbf{\textit{betas}} & \textbf{\textit{eps}} & \textbf{\textit{bs}} & \textbf{\textit{ws}} & \textbf{\textit{gss}} & \textbf{\textit{wp}} & \textbf{\textit{lr}} & \textbf{\textit{wd}} & \textbf{\textit{decay}} & \textbf{\textit{pr}} & \textbf{\textit{init.}} & \textbf{\textit{m.}}       \\ 
\hline\hline
\multicolumn{16}{|c|}{\textit{\textbf{Decoder}}}                                                                                                                                                                                                                                                                                                                                                                                                                                                      \\ 
\hline
GPT-C $366$M                               & {\color[rgb]{1,0,0}\texttimes}                      & BBPE                     & Adam                    & $-$                       & $-$                     & $-$                    & $1,024$                & $-$                     & $-$                    & $6.25$e-$5$              & $-$                    & Cosine                  & $-$                    & Scratch                & \textrightarrow                          \\ 
\hline
CodeGPT $124$M                             & {\color[rgb]{1,0,0}\texttimes}                      & BBPE                     & Adam                    & $-$                       & $-$                     & $-$                    & $768$                  & $-$                     & $-$                    & $5$e-$5$                 & $-$                    & $-$                       & $-$                    & GPT-2                  & \textrightarrow                          \\ 
\hline
GPT-Neo $2.7$B                             & $-$                      & BBPE                     & Adam                    & $0.9$,$0.95$                & $1$e-$8$                  & $-$                    & $2,048$                & $-$                     & $3,000$                & $-$                    & $0.1$                  & Cosine                  & $-$                    & Scratch                & \textrightarrow                          \\ 
\hline
GPT-J $6$B                                 & $-$                      & BBPE                     & Adam                    & $-$                       & $-$                     & $-$                    & $2,048$                & $16$                    & $3,000$                & $-$                    & $0.1$                  & $-$                       & BF16                 & $-$                      & \textrightarrow                          \\ 
\hline
Codex $12$B                                & {\color[rgb]{0,0.5,0}\checkmark}                      & BBPE                     & Adam                    & $0.9$,$0.95$                & $1$e-$8$                  & $2$M                   & $4,096$                & $-$                     & $175$                  & $1$e-$4$                 & $0.1$                  & Cosine                  & $-$                    & GPT-3                  & \textrightarrow                          \\ 
\hline
GPT-CC $1.3$B                          & {\color[rgb]{0,0.5,0}\checkmark}                      & BBPE                     & AdaFa                  & $-$                       & $-$                     & $-$                    & $1,024$                & $-$                     & $5,000$                & $2$e-$5$                 & $0.1$                  & Linear                  & $-$                    & GPT-Neo                & \textrightarrow                          \\ 
\hline
CodeParrot $1.5$B                          & {\color[rgb]{0,0.5,0}\checkmark}                      & BBPE                     & AdamW                   & $0.9$,$0.999$               & $1$e-$8$                  & $524$K                 & $1,024$                & $16$                    & $750$                  & $5$e-$5$                 & $0.1$                  & Cosine                  & $-$                    & Scratch                & \textrightarrow                          \\ 
\hline
LaMDA $137$B                               & $-$                      & SP                       & $-$                       & $-$                       & $-$                     & $256$K                 & $-$                    & $-$                     & $-$                    & $-$                    & $-$                    & $-$                       & $-$                    & $-$                      & \textrightarrow                          \\ 
\hline
PolyCoder $2.7$B                           & {\color[rgb]{0,0.5,0}\checkmark}                      & BBPE                     & AdamW                   & $0.9$,$0.999$               & $1$e-$8$                  & $262$K                 & $2,048$                & $-$                     & $1,600$                & $1.6$e-$4$               & $-$                    & Cosine                  & $-$                    & Scratch                & \textrightarrow                          \\ 
\hline
CodeGen $16.1$B                            & {\color[rgb]{0,0.5,0}\checkmark}                      & BBPE                     & Adam                    & $0.9$,$0.999$               & $1$e-$8$                  & $2$M                   & $2,048$                & $-$                     & $3,000$                & $0.5$e-$4$               & $0.1$                  & Cosine                  & $-$                    & $-$                      & \textrightarrow                          \\ 
\hline
InCoder $6.7$B                             & {\color[rgb]{0,0.5,0}\checkmark}                      & BBPE                     & Adam                    & $0.9$,$0.98$                & $-$                     & $-$                    & $2,048$                & $-$                     & $1,500$                & $-$                    & $-$                    & PN                      & $-$                    & Scratch                & $\leftrightarrow$                          \\ 
\hline
GPT-NeoX $20$B                             & {\color[rgb]{0,0.5,0}\checkmark}                      & BBPE                     & ZeRo                    & $0.9$,$0.95$                & $1$e-$8$                  & $3.15$M                & $2,048$                & $32$                    & $-$                    & $9.7$e-$5$               & $0.01$                 & Cosine                  & FP16                 & Scratch                & \textrightarrow                          \\ 
\hline
PaLM-Coder $540$B                          & {\color[rgb]{0,0.5,0}\checkmark}                      & SP                       & Adafa.                  & $-$                       & $-$                     & $-$                    & $2,048$                & $-$                     & $-$                    & $1$e-$2$                 & $-$                    & $-$                       & $-$                    & PaLM                   & \textrightarrow                          \\ 
\hline
PanGu-Coder $2.6$B                         & {\color[rgb]{0,0.5,0}\checkmark}                      & SP                       & Adam                    & $0.9$,$0.95$                & $-$                     & $-$                    & $1,024$                & $-$                     & $-$                    & $-$                    & $0.01$                 & Cosine                  & $-$                    & Scratch                & \textrightarrow                          \\ 
\hline
FIM $6.9$B                                                                                                        & {\color[rgb]{0,0.5,0}\checkmark}                      & BBPE                     & Adam                    & $-$                       & $-$                     & $2$M                   & $2,048$                & $-$                     & $-$                    & $2.4$e-$4$               & $-$                    & $-$                       & $-$                    & Scratch                & $\leftrightarrow$                          \\ 
\hline
PyCodeGPT $110$M                           & {\color[rgb]{0,0.5,0}\checkmark}                      & BBPE                     & AdamW                   & $0.9$,$0.95$                & $1$e-$8$                  & $480$K                 & $1,024$                & $4$                     & $1,000$                & $5$e-$4$                 & $0.1$                  & Cosine                  & FP16                 & Scratch                & \textrightarrow                          \\ 
\hline
CodeGeeX $13$B                             & $-$                      & BBPE                     & ZeRo                    & $0.9$,$0.95$                & $-$                     & $-$                    & $2,048$                & $-$                     & $-$                    & $-$                    & $0.1$                  & Cosine                  & FP16                 & $-$                      & \textrightarrow                          \\ 
\hline
BLOOM $176$B                               & {\color[rgb]{0,0.5,0}\checkmark}                      & BBPE                     & Adam                   & $0.9$,$0.95$               & $-$                  & $-$                 & $2,048$                & $-$                     & $-$                & $6$e-$5$               & $0.1$                    & Cosine                  & BF16                    & Scratch                & \textrightarrow                          \\ 
\hline
SantaCoder $1.1$B                          & {\color[rgb]{0,0.5,0}\checkmark}                      & BBPE                     & Adam                    & $0.9$,$0.95$                & $1$e-$8$                  & $-$                    & $-$                    & $-$                     & $-$                    & $2$e-$4$                 & $0.1$                  & Cosine                  & FP16                 & Scratch                & $\leftrightarrow$                          \\ 
\hline
\multicolumn{16}{|c|}{\textbf{\textit{Encoder-Decoder}}}                                                                                                                                                                                                                                                                                                                                                                                                                                              \\ 
\hline
PyMT5 $374$M                               & {\color[rgb]{0,0.5,0}\checkmark}                      & BBPE                     & Adam                    & $0.9$,$0.98$                & $1$e-$6$                  & $-$                    & $2,200$                & $-$                     & $5,000$                & $9.1875$e-$5$            & $0.01$                 & IS                      & FP16                 & $-$                      & \textrightarrow                          \\ 
\hline
PLBART $406$M                              & {\color[rgb]{1,0,0}\texttimes}                      & SP                       & Adam                    & $-$,$0.98$                  & $1$e-$6$                  & $-$                    & $768$                  & $-$                     & $-$                    & $5$e-$5$                 & $-$                    & Linear                  & FP16                 & $-$                      & \textrightarrow                          \\ 
\hline
CodeT5 $770$M                              & {\color[rgb]{1,0,0}\texttimes}                      & BBPE                     & AdamW                   & $-$                       & $-$                     & $-$                    & $-$                    & $-$                     & $1,000$                & $2$e-$4$                 & $0.05$                 & Linear                  & FP16                 & Scratch                & \textrightarrow                          \\ 
\hline
JuPyT5 $350$M                              & {\color[rgb]{0,0.5,0}\checkmark}                      & BBPE                     & Adam                    & $0.9$,$0.98$                & $1$e-$6$                  & $-$                    & $2,200$                & $-$                     & $5,000$                & $9.1875$e-$5$            & $0.01$                 & IS                      & FP16                 & PyMT5                  & \textrightarrow                          \\ 
\hline
AlphaCode~$41.1$B                          & {\color[rgb]{0,0.5,0}\checkmark}                      & SP                       & AdamW                   & $0.9$,$0.95$                & $-$                     & $-$                    & $6,144$                & $-$                     & $1,000$                & $1$e-$4$                 & $0.1$                  & Cosine                  & BF16                 & $-$                      & \textrightarrow                          \\ 
\hline
CodeRL $770$M                              & $-$                      & BBPE                     & AdamW                   & $-$                       & $-$                     & $-$                    & $-$                    & $-$                     & $-$                    & $-$                    & $-$                    & PN                      & $-$                    & CodeT5                 & \textrightarrow                          \\ 
\hline
CodeT5Mix $770$M                           & {\color[rgb]{0,0.5,0}\checkmark}                      & BBPE                     & AdamW                   & $-$                       & $-$                     & $-$                    & $-$                    & $-$                     & $-$                    & $-$                    & $0.1$                  & Linear                  & FP16                 & Scratch                & \textrightarrow                          \\ 
\hline
ERNIE-Code $560$M                          & {\color[rgb]{1,0,0}\texttimes}                      & SP                       & AdaFa                  & $-$                       & $-$                     & $-$                    & $1,024$                & $15$                    & $1,000$                & $1$e-$4$                 & $-$                    & Linear                  & BF16                 & mT5                    & \textrightarrow                          \\
\hline
\end{tabular}
}
}
\caption{The details of LLMs for \nlcode.
We list the full names of these abbreviations: de-duplication (\textbf{\textit{de.}}), tokenizer (\textbf{\textit{token.}}), optimizer (\textbf{\textit{opti.}}), batch size (\textbf{\textit{bs}}), window size (\textbf{\textit{ws}}), gradient accumulation steps (\textbf{\textit{gss}}), warmup steps (\textbf{\textit{wp}}), learning rate (\textbf{\textit{lr}}), weight decay (\textbf{\textit{wd}}), decay schedule (\textbf{\textit{decay}}), precision floating point (\textbf{\textit{pr}}), model initialization (\textbf{\textit{init.}}), left-to-right (\textrightarrow), fill-in-the-middle ($\leftrightarrow$), byte-level byte-pair-encoding (BBPE), SentencePiece (SP), polynomial (PN), and inverse square (IS).
}
\label{tab:models-details}
\end{table*}

\begin{table*}[t]
\centering
\scalebox{0.97}{
\rotatebox{0}{
\begin{tabular}{lll} 
\toprule
\textbf{Benchmark} & \textbf{Originate From}                                            & \multicolumn{1}{c}{\textbf{Multilingual}}                                                                                                                           \\ 
\toprule
MCoNaLa~\citeyearpar{mconala}            & CoNaLa~\citeyearpar{conala}                                                  & English, Spanish, Japanese, Russian                                                                                                                                 \\ 
\hline
ODEX~\citeyearpar{odex}               & \begin{tabular}[c]{@{}l@{}}CoNaLa~\citeyearpar{conala}\\MCoNaLa~\citeyearpar{mconala}\end{tabular} & English,~Spanish, Japanese, Russian                                                                                                                                 \\ 
\hline
MBXP~\citeyearpar{mbxp}               & MBPP~\citeyearpar{mbpp}                                                    & \begin{tabular}[c]{@{}l@{}}Python, Java, JavaScript, TypeScript, Go, Ruby,\\Kotlin, PHP, C\#, Scala, C++, Swift, Perl\end{tabular}                                  \\ 
\hline
MBXP-HumanEval~\citeyearpar{mbxp}     & HumanEval~\citeyearpar{codex}                                               & \begin{tabular}[c]{@{}l@{}}Python, Java, JavaScript, Ruby, Kotlin, PHP,~Scala,\\Swift, Perl,~\end{tabular}                                                          \\ 
\hline
MultiPL-MBPP~\citeyearpar{multipl-e}       & MBPP~\citeyearpar{mbpp}                                                    & \begin{tabular}[c]{@{}l@{}}Python, Java, JavaScrpt, TypeScript, Go, Ruby,\\Julia, PHP,~C\#, Scala, C++,~Swift, Perl, D, Bash,\\Racket,~Lua,~ R,~ Rust\end{tabular}  \\ 
\hline
MultiPL-HumanEval~\citeyearpar{multipl-e}  & HumanEval~\citeyearpar{codex}                                               & \begin{tabular}[c]{@{}l@{}}Python, Java, JavaScrpt, TypeScript, Go, Ruby,\\Julia, PHP,~C\#, Scala, C++,~Swift, Perl, D, Bash,\\Racket,~Lua,~ R,~ Rust\end{tabular}  \\ 
\hline
HumanEval-X~\citeyearpar{codegeex}        & HumanEval~\citeyearpar{codex}                                               & Python, Java, JavaScript, Go, C++                                                                                                                                   \\
\bottomrule
\end{tabular}
}
}
\caption{Details of multilingual \nlcode benchmarks. Here we also list MCoNaLa and CoNaLa, which have no test case for evaluation.}
\label{tab:benchmarks-multilingual}
\end{table*}

\end{document}